\documentclass[runningheads]{svmult}
\usepackage{makeidx}   
\usepackage{graphicx}  
\usepackage{subeqnar}  
\usepackage{multicol}  
\usepackage{physprbb}  
\newcommand{\newc}{\newcommand}
\newc{\beq}{\begin{equation}}
\newc{\eeq}{\end{equation}}
\newc{\barr}{\begin{eqnarray}}
\newc{\earr}{\end{eqnarray}}

\begin{document}
\title* { Neutrino properties studied with a triton source using
large TPC detectors}
\toctitle{ Neutrino properties \protect\newline studied with a
triton source using large TPC detectors}
%
%
\titlerunning{Direct SUSY Dark Matter Detection}
%
\author{Y. Giomataris$^{1}$ and J.D. Vergados$^{2}$ }
\authorrunning{Giomataris and Vergados}
%
%
\institute{CEA, Saclay, DAPNA, Gif-sur-Yvette, Cedex,France.
 \and
University of Ioannina, Ioannina, GR 45110, Greece.
\\E-mail:Vergados@cc.uoi.gr}
\maketitle              
\begin{abstract}
Abstract
The purpose of the present paper is to study the neutrino
properties as they may appear in the low energy neutrinos emitted
in triton decay:
$$^3_1 H \rightarrow ^3_2He +e^-+\tilde{\nu}_e$$
with maximum neutrino energy of $18.6~KeV$. The technical
challenges to this end can be summarized as building a very large
TPC capable of detecting low energy recoils, down to a few 100 eV,
within the required low background constraints. More specifically
We propose the development of a spherical gaseous TPC of about
10-m in radius and a 200 Mcurie triton source in the center of
curvature. One can list a number of exciting studies, concerning
fundamental physics issues, that could be made using a large
volume TPC and low energy antineutrinos: 1) The oscillation length
involving the small angle $\delta=\sin{2 \theta_{13}}$ in the
$\nu_e$ disappearance experiment is comparable to the length of
the detector. Measuring the counting rate of neutrino-electron
elastic scattering as function of the distance of the source will
give a precise and unambiguous measurement of the oscillation
parameters free of systematic errors. In fact first estimations
show that a sensitivity of a few percent for the measurement of
the above angle. 2) The low energy detection threshold offers a
unique sensitivity for the neutrino magnetic moment which is about
two orders of magnitude beyond the current experimental limit of
$10^{-10}\mu_B$. 3) Scattering at such low neutrino energies has
never been studied and any departure from the expected behavior
may be an indication of new physics beyond the standard model.
 We present a summary of various theoretical
studies and possible measurements.
\end{abstract}
\section{Introduction.}
\label{secint}

Neutrinos are the only particles in nature, which are
characterized by week interactions only. They are thus expected to
provide the laboratory for understanding the fundamental laws of
nature. Furthermore they are electrically neutral particles
characterized by a very small mass. Thus it is an open question
whether they are truly neutral, in which case the particle
coincides with its own antiparticle , i.e. they are Majorana
particles, or they are characterized by some charge, in which case
they are of the Dirac type, i.e the particle its different from
its antiparticles \cite{VERGADOS}. It is also expected that the
neutrinos produced in week interactions are not eigenstates of the
world Hamiltonian, they are not stationary states, in which case
one expects them to exhibit oscillations \cite{VERGADOS,VOGBEAC} .
As a matter of fact such neutrino oscillations seem to have
observed in atmospheric neutrino \cite{SUPERKAMIOKANDE},
interpreted as
 $\nu_{\mu} \rightarrow \nu_{\tau}$ oscillations, as well as
 $\nu_e$ disappearance in solar neutrinos \cite{SOLAROSC}. These
 results have been recently confirmed by the KamLAND experiment \cite{KAMLAND},
 which exhibits evidence for reactor antineutrino disappearance.
 This has been followed by an avalanche of interesting analyses
 \cite{BAHCALL02}-\cite{BARGER02}.
 The purpose of the present paper is to discuss a new experiment to
study the above neutrino properties as they may appear in the low
energy neutrinos emitted in triton decay:
$$^3_1 H \rightarrow ^3_2He +e^-+\tilde{\nu}_e$$
with maximum neutrino energy of $18.6~KeV$. The detection will be
accomplished employing gaseous Micromegas,large TPC detectors with
good energy resolution and low background \cite{GIOMATAR}. In
addition in this new experiment we hope to observe or set much
more stringent constraints on the neutrino magnetic moments. This
question has have been very interesting for a number of years and
it has been revived recently \cite{VogEng}-\cite{TROFIMOV}. The
existence of the neutrino magnetic moment can be demonstrated
either in neutrino oscillations in the presence of strong magnetic
fields or in electron neutrino scattering. The latter is expected
to dominate over the weak interaction in the triton experiment
since the energy of the outgoing electron is very small.
Furthermore the possibility of directional experiments will
provide additional interesting signatures. Even experiments
involving polarized electron targets are beginning to be
contemplated \cite{RASHBA}. There are a number of exciting
studies, of fundamental physics issues, that could be made using a
large volume TPC and low energy antineutrinos:
\begin{itemize}
\item The oscillation length is comparable to the length of the detector. Measuring
the counting rate of neutrino elastic scattering as function of
the distance of the source will give a precise and unambiguous
measurement of the oscillation parameters free of systematic
errors. First estimations show that a sensitivity of a few percent
for the measurement of $\sin^2{\theta_{13}}$.
\item The low energy detection threshold offers a unique sensitivity for the
neutrino magnetic moment, which is about two orders of magnitude
beyond the current experimental limit of $10^{-10}\mu_B$. In our
estimates below we will use the optimistic value of
$10^{-12}\mu_B$.
\item Scattering at such low neutrino energies has never been studied before. In
addition one may exploit the extra signature provided by the
photon in radiative electron neutrino scattering. As a result any
departure from the expected behavior may be  an indication of
physics beyond the standard model.
\end{itemize}
 In the following we will present a summary of various theoretical studies and
possible novel measurements

\section{The neutrino mixing.}
\label{secnm}
 We suppose that the neutrinos produced in week
interactions are not stationary states, i.e. they are not
eigenstates of the Hamiltonian. In this case the week eigenstates
are linear combinations of the mass eigenstates \cite{VERGADOS}:
\begin{equation}
\nu^{0}_{e L}=\sum^3_{k=1} ~U^{(11)}_{ek}~\nu_{kL} +
\sum^3_{k=1} ~U^{(12)}_{ek}~N_{kL},
\label{eq:1.2}
\end{equation}
\begin{equation}
\nu^{0}_{e R}=\sum^3_{k=1} ~U^{(21)}_{ek}~\nu_{kL} +
\sum^3_{k=1} ~U^{(22)}_{ek}~N_{kL},
\label{eq:1.3}
\end{equation}
the fields
$$\nu_{k}~(N_{k})$$
are the light (heavy) Majorana neutrino eigenfields with masses\\
$m_k~$ ($m_k << 1$ MeV) and $M_k$ ($M_k >> 1$ GeV).

The matrices
  $$U^{(11)}_{ek}~~,~~ U^{(22)}_{ek}$$
 are approximately unitary, while the matrices
$$U^{(12)}_{ek}~~,~~ U^{(21)}_{ek}$$
 are very small (of order of mass of the up
quark  divided by that of the heavy neutrino, $m_N\approx 10^{12}~GeV$)\\
$\nu_k,N_k$ satisfy the Majorana condition:
$$\nu_k \xi_k = C ~{\overline{\nu}}_k^T~,~N_k \Xi_k = C ~{\overline{N}}_k^T$$
where C denotes the charge conjugation. The quantities:
 $$\xi_k=e^{i\lambda_k}~, ~\Xi_k=e^{i\Lambda_k}$$
are phase factors, which guarantee that the eigenmasses are positive. They are relevant
even in a CP conserving theory since even then some of the phases
 $\lambda_k~, ~\Lambda_k$ can take the value $\pi$.

In what follows we will ignore the heavy neutrino components, i.e. we will assume that
$U=U^{11}$.
\section{Neutrino masses as extracted from various experiments}
\label{secnme}
 At this point it instructive to elaborate a little
bit on the neutrino mass combinations entering various
experiments.
\begin{itemize}
\item Neutrino oscillations. \\
These in principle, determine the  mixing matrix and
   two independent mass-squared differences, e.g.
 $$\Delta m^2_{21} = m^2_2 - m^2_1~~,~~ \Delta m^2_{31} = m^2_3 - m^2_1$$
They cannot determine:
\begin{enumerate}
\item the scale of the masses, e.g. the lowest
eigenvalue $m_1$ and
\item the two relative Majorana phases.
\end{enumerate}
\item The end point triton decay.

This can determine one of the masses, e.g. $m_1$ by measuring:
\begin{equation}
(m_{\nu})_{1 \beta} \equiv m_{\nu}=| \sum_{j=1}^{3}
 U^*_{ej}U_{ej} m^2_j |^{1/2} \, , ~U=U^{11}
\label{mass.3}
\end{equation}
 Once $m_1$ is known one can find\\
  $$m_2=[\delta m^2_{21}+m^2_1]^{1/2}~~,~~m_3=[\delta m^2_{31}+m^2_1]^{1/2}$$
 provided, of course that the mixing matrix is known. \\
Since the Majorana phases do not appear, this experiment cannot
differentiate between Dirac and Majorana neutrinos. This can only
be done via lepton violating processes, like: \item $0\nu \beta
\beta $ decay.

This provides an additional   independent linear combinations of
the masses and the Majorana phases.
\begin{equation}
\langle m_{\nu}\rangle_{2 \beta} \equiv \langle m_{\nu}\rangle =
|\sum _{j=1}^{3} U_{ej}U_{ej}e^{i\lambda _j} m_j| \label{mass.1}
\end{equation}
\item and muon to positron conversion.

This also provides an additional relation
\begin{equation}
\langle m_{\nu}\rangle_{\mu e+}=|\sum_{j=1}^{3}
                   U^*_{\mu j}U^*_{ej}e^{-i\lambda _j}m_j| \, .
\label{mass.2}
\end{equation}
\end{itemize}
 Thus the two independent relative CP phases can in principle be measurable.
{\bf So these three types of experiments together can specify all
parameters not settled by the neutrino oscillation experiments}.

 Anyway from the neutrino oscillation data alone we cannot infer
 the mass scale. Thus the following scenarios emerge
 \begin{enumerate}
 \item the lightest neutrino is $m_1$ and its mass is very small.
 This is the normal hierarchy scenario. Then:
$$\Delta m^2_{21} = m^2_2~~,~~ \Delta m^2_{31} = m^2_3$$
\item The inverted hierarchy scenario. In this case the mass $m_3$
is very small. Then:
$$\Delta m^2_{21} = m^2_2 - m^2_1~~,~~ \Delta m^2_{31} = m^2_1$$
\item The degenerate scenario. In such a situation all masses are about
equal and much larger than the differences appearing in neutrino
oscillations. In this case we can obtain limits on the mass scale
as follows:
\begin {itemize}
\item From  triton decay. Then \cite{LOBASHEV}
$$ m_1 \approx (m_{\nu})_{1 \beta}\leq 2.2 eV$$
This limit is expected to substantially improve in the future
\cite{KATRIN}.
\item From  $0\nu~\beta\beta$ decay. The analysis now depends on the mixing
matrix and the CP phases of the Majorana neutrino eigenstates
\cite{VERGADOS} (see discussion below).  If the relative phase of
the CP eigenvalues of the two strongly admixed states is zero, the
best limit coming from $0\nu~\beta\beta$ decay is:
 $$m_1 \approx\langle m_{\nu}\rangle_{2 \beta}\leq 0.5~eV$$
 On the other hand it is
$$m_1 \approx \frac{\langle
m_{\nu}\rangle_{2 \beta}}{\cos{2 \theta_{solar}}}\approx 2 \langle
m_{\nu}\rangle_{2 \beta}\leq 1.0~eV,$$
 if this relative phase is $\pi$.

These limits are going to greatly improve in the next generation
of experiments, see e.g. the review by Vergados  \cite{VERGADOS}
and the experimental references therein.
\end{itemize}
\end{enumerate}
\section{Experimental considerations}
\label{secec}
 In this section we will focus on the experimental
considerations
\subsection {The radial TPC concept}

\begin{figure}
\hspace*{-0.0cm}
\includegraphics[height=.4\textheight]{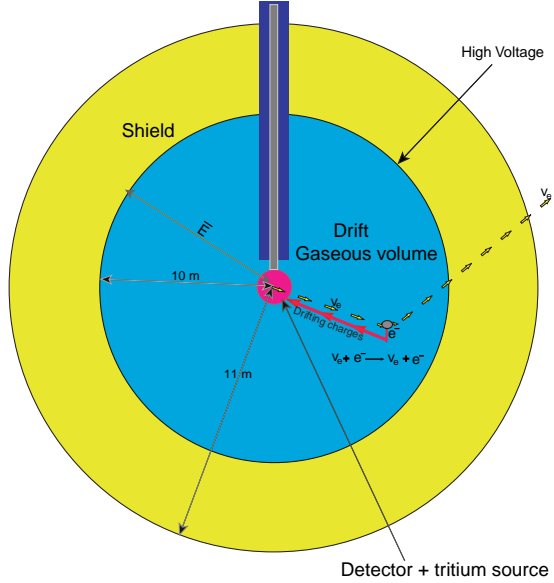}
\caption{ The principal features of the proposed TPC.
 \label{draw1} }
\end{figure}

One of the attractive features of the gaseous TPC is its ability
to precisely reconstruct particle trajectories without precedent
in the  redundancy of experimental points, i.e. a bubble chamber
quality with higher accuracy and real time recording of the
events. Many proposals are actually under investigation to exploit
the TPC advantages for various astroparticle projects and
especially solar or reactor neutrino detection and dark matter
search \cite{GORO}-\cite{SNOWDEN}. A common goal is to fully
reconstruct the direction of the recoil particle trajectory, which
together with energy determination provide a valuable piece of
information. The virtue of using the TPC concept in such
investigations has been now widely recognized and a special
International Workshop has been recently organized in Paris
\cite{PARIS}. The study of low energy elastic neutrino-electron
scattering using a strong tritium source was envisaged in by
Bouchez and Giomataris \cite{BG} employing a large volume gaseous
cylindrical TPC. We will present here an alternate detector
concept with different experimental strategy based on a spherical
TPC design. A sketch of the principal features of the proposed TPC
is shown in Fig. \ref{draw1}.

\begin{figure}
\hspace*{-0.0 cm}
\includegraphics[height=.4\textheight]{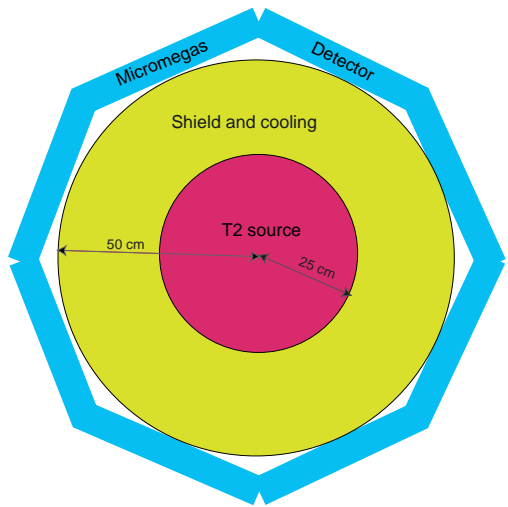}
\caption{ A schematic view of the inner part of the vessel with
the detector and the tritium source.
 \label{draw2} }
\end{figure}

 It consists of a spherical vessel of 10 meters in radius that contains about 20 Tons of
gas made by a thin low radioactivity acrylic metalized in the
inner surface to provide electric contact to the high voltage
(drift voltage of about -100 kV). It is located in a cavern of an
underground laboratory and separated from the rock by 1 meter
shield (2-3 meters water equivalent of high purity shielding). The
ground plane is another smaller sphere about 50 cm in radius which
carries the detector plane and defines with the drift volume a gas
target volume 9.5 m long; ionization electrons released during the
elastic scattering with the target gas are drifting to the
detector where are collected and amplified. There are actually
many gaseous detectors adequate for this experiment but we will
focus our detection strategy on Micromegas \cite{GIOMATAR}, a new
technology, which is now widely recognized and  used by many
particle physics experiments. The 200 Mcurie tritium source
container is a sphere 20-cm in radius. Neutrino emitted can
produce electron recoil in the gaseous volume by the elastic
scattering reaction and the distance from the center of curvature
is detected. The concept of the spherical TPC simplifies the whole
structure since:
\begin{itemize}
\item It provides in a simple way a good estimation of the radial location (depth) of
the neutrino-electron elastic scattering
\item It does not require a special field cage as in the conventional-cylindrical TPC.
\item the converging radial electric field is strongly focussing the drifting electron
charges providing a reasonable size of the detection plane (about
$4 m^2$)
\item Last, but not least, the whole structure is relatively simple and cheap
with a very-reasonable number of detectors and electronics
\end{itemize}
 A schematic view of the inner part vessel with the detector and the
tritium source is shown in Fig. \ref{draw2}.

Our approach is radically different from all other neutrino
oscillation experiments in that it measures the neutrino
interactions, as a function of the distance source-interaction
point, with an oscillation length that is fully contained in the
detector; it is equivalent to many experiments made in the
conventional way where the neutrino flux is measured in a single
space point.
  Furthermore, since the oscillation length is comparable to the
  detector depth, we expect an
exceptional signature: a counting rate oscillating from the triton
source location to the depth of the gas volume, i.e. at first a
decrease, then a minimum and finally an increase. In other words
we will have a full observation of the oscillation process as it
has already been done in accelerator experiments with neutral
strange particles ($K^0$).

\subsection {The gas vessel}

The external drift-electrode sphere will be made using low
background materials. Polypropylene or polyethylene have the
advantage that they do not contain oxygen in the bulk material.
The drift sphere could be enclosed in an external pressure vessel
containing, for instance, a low-radioactivity appropriate solid or
liquid that is used as a buffer shield to cope with the rock
background emission. Indeed the various background contributions
evaluated by other groups show that natural radioactivity of the
rock turns out to be by far the most important background
component in such investigations. The gas volume acts both as
target and detector of neutrino-electron elastic scattering
events. The total elastic scattering cross section for the triton
neutrinos is $0.58 \times 10^{-47} cm^2$ integrating between 100eV
and 1.27 keV, the maximum electron recoil energy. Filling the gas
volume with Xenon gas at atmospheric pressure ( containing about
$6\times 10^{30}$ electrons) will allow observation of 3500
elastic interactions/year using a 200 MCurie, 20 kg tritium
source. The versatility of the gas volume – Micromegas detector
scheme allows a variety of gaseous targets: Xenon can be used at
atmospheric pressure and has not any intrinsic radioactivity. It
contains, however, a large fraction of 85Kr that is a beta emitter
with end point energy of 700 keV and a quite short time of life of
10.7 years. However only a small part of the spectrum, about
$0.1\%$, is contained in our energy bandwidth. First estimates
show that a removal of Krypton at the level of $10^{-12}$ is
required. The later requirement is about 3 orders of magnitude
lower than in the liquid Xenon future WIMP searches projects
\cite{ICHIGE,LUSCHER}. Since, however, Xenon is the most expensive
noble gas other gas targets should be considered. Argon is
very-cheap material and must operate at a pressure of about 2.5
bars. It has some intrinsic radioactivity mainly due to a beta
emitting isotope $^{42}Ar$ ($t_{1/2}=33~y$, $E_{max}=565~ keV$).
The Icarus group \cite{ICARUS} has measured the ratio
$^{42}Ar/^{38}Ar$ to be less than $7\times 10^{-22}$; our first
preliminary estimations show that the effect of this radioactivity
is quite small and thus Argon should not be excluded as target for
this experiment. Neon has no intrinsic radioactivity and because
of its low boiling point is easy to purify and clean for unwanted
impurities, but it must be used at a pressure of 5 bars. Helium is
low cost and the cleanest gaseous target, but it has the drawback
that its density is quite low , $166gr/m^3$ at NTP, which implies
that it should operate at a pressure of about 25 bars.

The TPC is located underground and it is fully enclosed by the
drift spherical electrode, which at the same time constitutes an
efficient Faraday cage. We can then reasonably assume that the
noise seen by the TPC will be generated only by its internal
components. The radio purity of the various elements is one of the
main challenges and, to deal with it, we need other groups
participating in this project, in particular those who  are world
leaders in ultraslow background technology as well.

\subsection {The detector}

The energy released by ionizing particles (low energy electron
recoils) in the drift volume will produce local charge clouds to
be transported to the detector plane in order to be amplified and
collected on the anodes.  Since we are dealing with low energy
recoils we need a high gain detector with good time resolution and
capable to reject other backgrounds induced by Compton scattering
of gamma rays, beta decays or other ionizing particles, such as
small contaminants of radon, present in underground facilities. To
meet the various objectives we will concentrate on Micromegas
technology. The European experiment to search for solar axions
\cite{ZIOUTAS,AALSETH}, CAST, is using the Micromegas idea, and
several successful experiments have also been using such charge
readout \cite{MAGNON,ADRIAN}. MICROMEGAS (MICROMEsh GAseous
Structure) \cite{GIOMA98,DERRE99,DERRE01} is a gaseous two - stage
parallel-plate avalanche chamber design consisting of a narrow
$50-100 \mu m$ amplification gap and a thick conversion region,
separated by a light conducting micromesh, usually made of
electroformed nickel or copper. Electrons released in the
gas-filled conversion gap by an ionizing particle are transported
to the amplification gap where they are multiplied in an avalanche
process. In most of the applications signals induced on anode
elements (strips or pads) are providing a precise x-y spatial
projection of the energy deposition that is a key element for
efficient background rejection. For this project such precise two-
dimensional determination is not required and will be only
optional in the cases for which  the background level is so high
that additional rejection will be needed. The detector element has
a hexagonal-flat shape with a dimension of about 20- cm. A lot of
such detector elements, about $100$, are arranged around a
spherical surface. Each detector has a single read-out that
consists of a low noise charge preamplifier (about 50 ns peaking
time) followed by fast shaper and a 100 MHz flash ADC. Building
such Micromegas detectors do not present a major technological
effort since that size counters are routinely used in particle
beams; larger detectors of $40 \times40 cm^2$ are nicely operating
in COMPASS experiment. Detecting such low energy recoils with
Micromegas detectors at NTP is not a big deal. At high pressure
there is a certain drop of the gain which is proportional to the
value of the pressure. In the later case we could rely on various
future developments and, in this context, we should point out the
progress made in the framework of the HELLAZ \cite{GORO}
experiment. An exhaustive study, made for solar neutrino
detection, using high pressure helium has shown the ability of the
Micromegas detector to reach high gains (about $10^6$) at $20$~
bars, which opens the possibility to lower the detection threshold
to a single electron and therefore the single electron counting.

To summarize:
\begin{itemize}
\item The aim of the proposed detector will be the detection of very low
 energy neutrinos emitted by a strong tritium source through their elastic
 scattering on electrons of the target.
\item The $(\nu,e)$ elastic differential cross section is the sum of the
 charged and neutral current contributions (see sec. \ref{secens}) and
is a function of the energy. It is, however,  it is quite small,
 see Eq. (\ref{weekval}).
\item Integrating this cross section up to energies of $15~KeV$ we get a very
 small value, $\sigma=0.4\times 10^{-47}cm^2$. This means that, to get a
significant signal in the detector, for 200 Mcurie
tritium source (see next section) we will need about 20 kilotons of gaseous
 material.
\item The elastic $(\nu,e)$ cross section, being dominated by the charged
 current, especially for low energy electrons (see Fig. \ref{chi} below),
will be different from that of the other flavors, which is due to the
 neutral current alone. This will allow us to observe neutrino oscillation
enabling a modulation on the counting rate along the oscillation length.
The effect depends on the electron energy T as is shown in sec. \ref{secno}
\end{itemize}
\subsection {The neutrino source}

Tritium is widely produced in nuclear reactors using light water
and especially in those using heavy water, the back-up production
technology being the Linear accelerator where tritium is usually
made by capturing neutrons in $^3_2He$ (helium gas). Tritium has a
relatively short half-life of 12.3 years, which is long enough to
ensure a high neutrino flux for several years of experimental
investigations. It emits a low energy beta particle (energy of
about 5 keV), and an anti-neutrino (energy of about 5 keV ) and in
the process decays to Helium-3 which is not radioactive. Absorbing
such low energy beta particles is not a big deal, a few millimeter
copper sheet will stop the total emitted beta energy or soft
X-rays from bremsstrahlung process. The total power produced is 4
kwatt/20 kg that must be dissipated by an appropriate cooling
system; a liquid circulating system in the volume surrounding the
source  must be designed. Temperature measurement of the heating
loss will provide the neutrino flux to within one percent. Large
amount of tritium radioactive material ($ >>20~ kgr$) are stored
in various parts of the world due to  the reduction of nuclear
weapons or production by nuclear reactors  (in particular those
using heavy water).

The container of the source should be carefully designed in order
to fulfill the safety requirements and, at the same time, provide
a flexible moving system for source on- off measurements. It could
be made, for instance, of low radioactivity copper about 1 cm
thick. The space between the source and the detector plane is
filled by a high radioactive purity material which must ensure, at
the same time, adequate cooling to compensate for the power
produced by the triton emission energy. The design, construction,
test and transport of the source system to the underground
laboratory is certainly a delicate project that requires a team of
specialized physicists and engineers. We would like also to
mention a more exhaustive study of such intense tritium source
made by another group \cite{NEGANOV}.

\subsection {Simulation and results}

We assume a spherical type detector, described in the previous
section, filed with Xenon gas at NTP and a tritium source of 20
kg, providing a very-high intensity neutrino emission of $6
\times10^{18} /s$. The Monte Carlo program is simulating all the
relevant processes:
\begin{itemize}
\item Beta decay and neutrino energy random generation
\item Oscillation process of $\nu_e$ due to the small mixing $\theta_{13}$ (see
Eq. \ref{osceq} below).
\item Neutrino elastic scattering with electrons of the gas target
\item Energy deposition, ionization processes and transport of charges to the
Micromegas detector.
\end{itemize}
The collected charge on the detector will provide the electron
recoil energy with a good precision. The lack of trigger signal,
however, will not allow a direct measurement of the radial
distance in the conventional way used in drift chambers. We have
adopted  a novel method of estimating the radial dimension, which
relies on the excellent time resolution of the detector (below 1
ns has been achieved \cite{GIOMA98}) and its ability to detect low
amount of charges (down to single electrons). The idea is to
exploit the large longitudinal diffusion of charges, produced by
energy deposition of the recoil electron, during their long drift
to the detector plane. The special electric configuration with a
very weak value ($E=10~Volt/cm$) at the highest distance (at 10 m)
works towards this goal; the longitudinal diffusion at such low
field is roughly proportional to E and the drift velocity
inversely proportional to E, enhancing time dispersion of the
collected signals. Measuring the arrival time of the ionization
electrons and therefore their time dispersion will provide a rough
but good estimation of the radial drift distance (L).

\begin{figure}
\hspace*{-0.0 cm}
\includegraphics[height=.4\textheight]{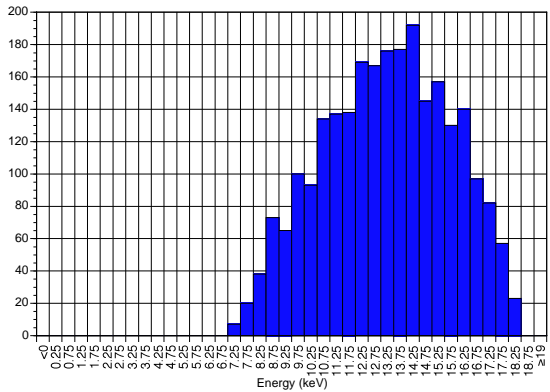}
\caption{ Neutrino energy distribution with an energy cutoff of
$200~eV$.
 \label{ioa3} }
\end{figure}

 First Monte
Carlo simulate are giving a resolution of better than 10 cm, which
is good enough for our need. In Fig. \ref{ioa3} the energy
distribution of the detected neutrinos, assuming a detection
threshold of 200 eV, is exhibited. The energy is concentrated
around 13 keV with a small tail to lower values.

In Figs \ref {osc} and \ref{osc2} below we  show the number of
detected elastic events as function of the distance L in bins of
one meter for several hypothesis for the value of the mixing angle
$\theta^2_{13}= 0.170, 0.085$ and $0.045$.
 We observe a decreasing of the signal up to about
6.5 m and then a rise. Backgrounds are not yet included in this
simulation but the result looks quite promising; even in the case
of the lowest mixing angle the oscillation is seen, despite
statistical fluctuations. We should point out that in the context
of this experiment complete elimination of the backgrounds is not
necessary. It is worth noting that:
\begin{itemize}
\item A source-off measurement at the beginning of the experiment will yield the
background level to be subtracted from the signal.
\item Fitting the observed oscillation pattern will provide, for the first time, a stand
alone measurement of the oscillation parameters, the mixing angle
and the square mass square difference.
\item Systematic effects due to backgrounds or to bad estimates of the neutrino
flux, which is the main worry in most of the neutrino experiments,
are highly reduced in this experiment.
\end{itemize}

\section{ A simple phenomenological neutrino mixing matrix}
\label{secmat}
 The available neutrino oscillation data (solar \cite{SOLAROSC} and
  atmospheric \cite{SUPERKAMIOKANDE})as well as the
 KamLand \cite{KAMLAND} results can
adequately be described by the following matrix:

$\left ( \begin{array}{c}\nu_e \\
 \nu_{\mu} \\
\nu_{\tau} \end{array} \right )=
\left ( \begin{array}{ccc}
c&s&\delta\\
-\frac{s+c \delta}{\sqrt 2}&  \frac{c-s \delta}{\sqrt 2}& \frac{1}{\sqrt 2}\\
 \frac{s-c \delta}{\sqrt 2}& -\frac{c+s \delta}{\sqrt 2}& \frac{1}{\sqrt 2}\\
 \end{array} \right )=
\left ( \begin{array}{c}\nu_1 \\
\nu_2 \\
\nu_3 \end{array} \right )$\\

Up to order $\delta^2$ ($\delta^2=4 \times 10^{-2}$). Sometimes we
will use $\theta_{13}$ instead of $\delta$. In the above
expressions
$$c=\cos \theta_{solar}~,~s=\sin \theta_{solar}$$
This angle is determined from the solar neutrino data
\cite{SOLAROSC}, \cite{BAHCALL02}-\cite{BARGER02}
$$\tan^2 \theta_{solar} \approx  0.35-0.42$$
while the analysis of KamLAND results
\cite{BAHCALL02}-\cite{BARGER02} yields:
$$\tan^2 \theta_{solar} \approx  0.64-0.79$$

\section{ Simple expressions for neutrino oscillations}
\label{secno}
\begin{itemize}
\item Solar neutrino Oscillation (LMA solution) is given by:
$$P(\nu_e \rightarrow \nu_e)\approx 1-(\sin 2 \theta_{solar})^2 \sin^2(\pi \frac{L}{L_{21}})$$
$$L_{21}=\frac{4 \pi E_{\nu}}{\Delta m^2_{21}}$$
The analysis of both the neutrino oscillation experiments as well
as KamLAND  \cite{BAHCALL02}-\cite{BARGER02} yield
$$\Delta
m^2_{21}=|m_2^2-m_1^2|=(5.0-7.5)\times 10^{-5}(eV)^2$$
\item The
Atmospheric Neutrino Oscillation takes the form:
$$P(\nu_{\mu} \rightarrow \nu_{\tau})\approx 2(\cos \theta_{solar})^2 \sin^2(\pi \frac{L}{L_{32}})$$
$$L_{32}=\frac{4 \pi E_{\nu}}{\Delta m^2_{32}} \rightarrow \Delta m^2_{32}=|m_3^2-m_2^2|=2.5\times 10^{-3}(eV)^2$$
\item We conventionally write
$$\Delta m^2_{32}=\Delta m_{atm}^2~~,~~\Delta m^2_{21}=\Delta m_{sol}^2$$
\item Corrections to disappearance experiments
\beq P(\nu_e \rightarrow \nu_e)= 1-\frac{(\sin 2 \theta_{solar})^2
\sin^2(\pi \frac{L}{L_{21}}) +4 \delta^2 \sin^2(\pi
\frac{L}{L_{32}})}{(1+\delta^2)^2}
 \label{osceq}
 \eeq
 \item The probability for $\nu_e\rightarrow\nu_{\mu}$ oscillation
 takes the form:
\beq P(\nu_e \rightarrow \nu_{\mu})= \frac{ \left [(\sin 2
\theta_{solar})^2 +\delta \sin{4\theta_{solar}}\right] \sin^2(\pi
\frac{L}{L_{21}}) +4 \delta^2 \sin^2(\pi
\frac{L}{L_{32}})}{(1+\delta^2)^2}
 \label{osceqa}
 \eeq
\item While the oscillation probability  $\nu_e\rightarrow\nu_{\tau}$ becomes:
 \beq P(\nu_e \rightarrow \nu_{\tau})= \frac{ \left [(\sin 2
\theta_{solar})^2 -\delta \sin{4\theta_{solar}}\right] \sin^2(\pi
\frac{L}{L_{21}}) +4 \delta^2 \sin^2(\pi
\frac{L}{L_{32}})}{(1+\delta^2)^2}
 \label{osceqb}
 \eeq
 \end{itemize}
  From the above expression we see that the small
amplitude $\delta$ term dominates in the case of triton neutrinos
($L \le L_{32}~,L_{21}=50L_{32}$)

\begin{figure}
\hspace*{-0.0 cm}
\includegraphics[height=.2\textheight]{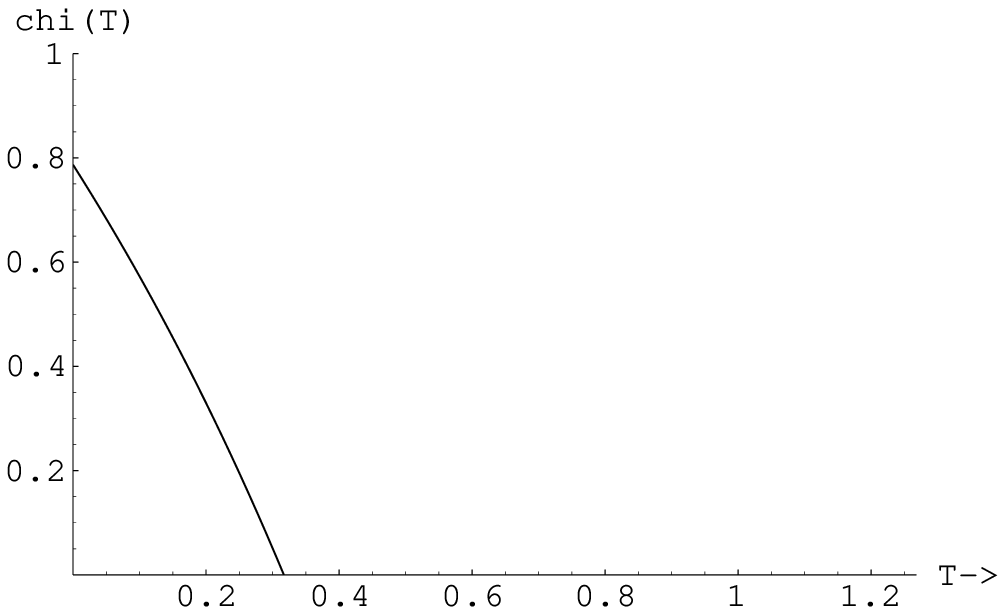}
\includegraphics[height=.2\textheight]{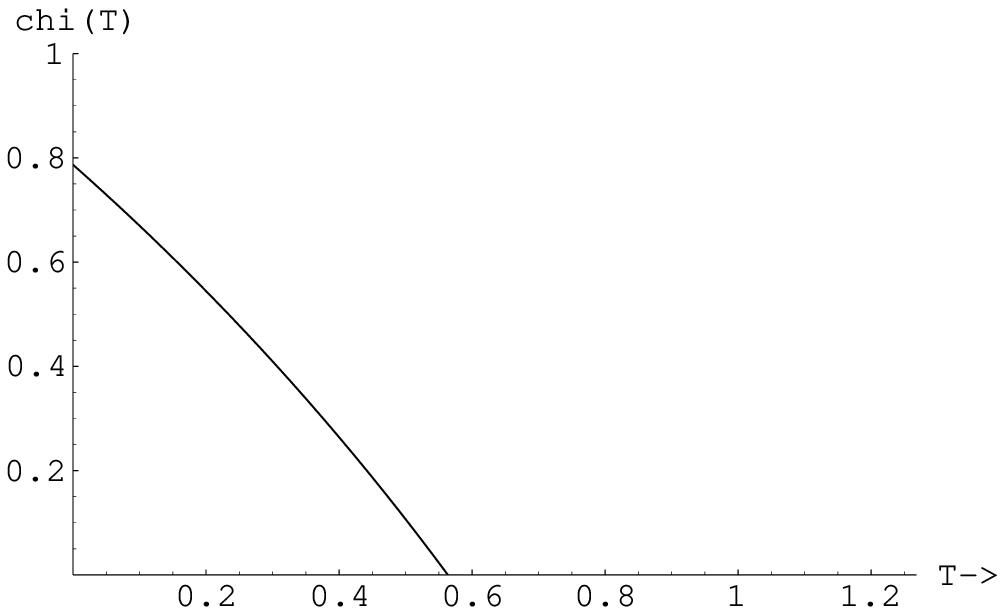}
\caption{ The parameter $\chi$ as a function of the electron
kinetic energy T for $E_{\nu}=9.0~KeV$ on the left and $12.0~KeV$
on the right.
 \label{chi1} }
\end{figure}

\begin{figure}
\hspace*{-0.0 cm}
\includegraphics[height=.2\textheight]{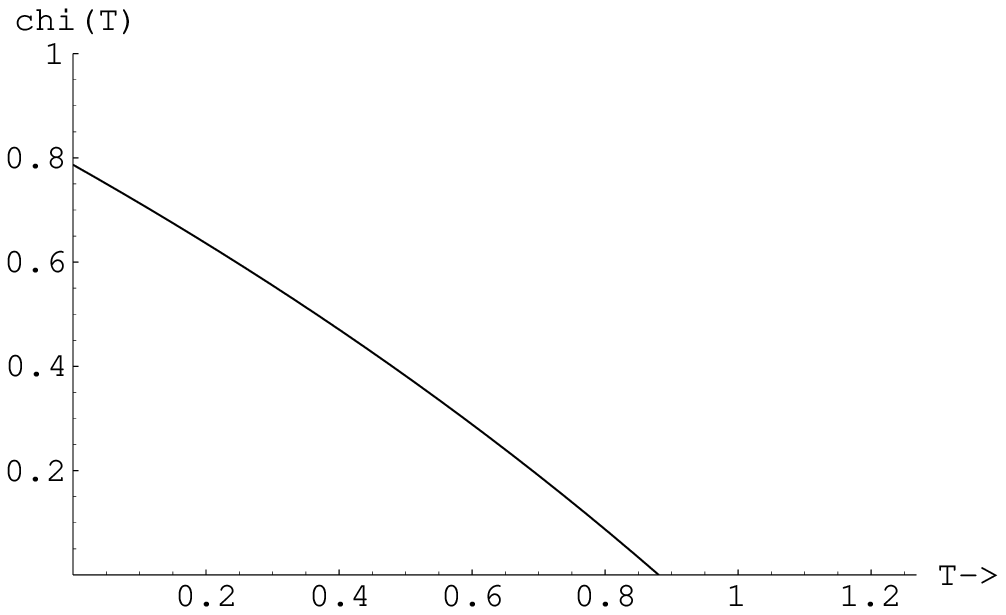}
\includegraphics[height=.2\textheight]{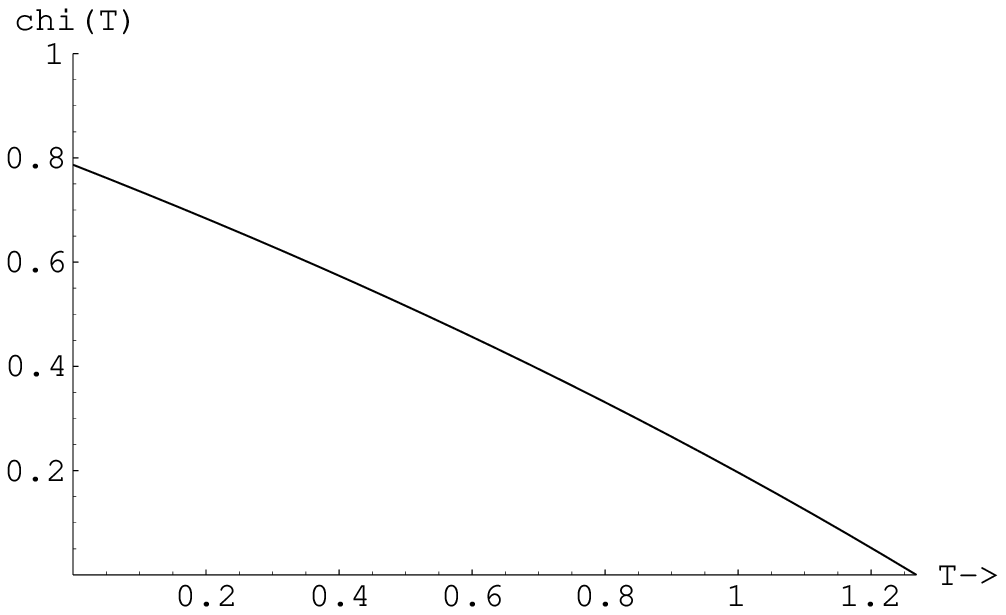}
\caption{ The parameter $\chi$ as a function of the electron
kinetic energy T for $E_{\nu}=15.0~KeV$ on the left and $18.0~KeV$
on the right.
 \label{chi2} }
\end{figure}

 In the proposed experiment the neutrinos will be detected via the
 recoiling electrons. If the neutrino-electron cross section were the same for
 all neutrino species one would not observe any oscillation at
 all. We know, however, that the electron neutrinos behave very
 differently due to the charged current contribution, which is not
 present in the other neutrino flavors. Thus the number of the
 observed electron events ($ELEV$) will vary as a function of $L/E_{\nu}$ as
 follows:
 \beq
ELEV \propto \left [1-\chi(E_{\nu},T) \frac{(\sin 2
\theta_{solar})^2 \sin^2(\pi \frac{L}{L_{21}}) +4 \delta^2
\sin^2(\pi \frac{L}{L_{32}})}{(1+\delta^2)^2} \right
]\frac{d(\sigma(\nu_e,e^-))}{dT}
 \label{eventeq}
 \eeq
where
$$\chi(E_{\nu},T)=\frac{(d\sigma(\nu_e,e^-))/dT-d(\sigma(\nu_{\alpha},e^-))/dT}{d(\sigma(\nu_e,e^-))/dT}$$
($\nu_{\alpha}$  is either $\nu_{\mu}$ or $\nu_{\tau}$). In other
words $\chi$ represents the fraction of the $\nu_e$-electron
cross-section, $\sigma(\nu_e,e^-)$, which is not due to the
neutral current. Thus the apparent disappearance oscillation
probability will be quenched by this fraction. As we will see
below, see section \ref{secens}, the parameter $\chi$, for
$sin^2\theta_W=0.2319$, can be cast in  the form: \beq
\chi(E_{\nu},T)=2\frac{2-(m_eT/E^2_{\nu})}{4.6199+0.4638(1-T/E{\nu})^2-1.4638(m_eT/E^2_{\nu})}
 \label{chi}
 \eeq
We thus see that the parameter $\chi$ depends not only on the
neutrino energy, but on the electron energy as well, see Figs
\ref{chi1}-\ref{chi2}.
It interesting to see that, for a given neutrino energy, $\chi$,
as a function of $T$, is almost a straight line. We notice that,
for large values of $T$, the factor $\chi$ is suppressed, which is
another way of saying that, in this regime, in the case of
$(\nu_e,e^-)$ differential cross-section the charged current
contribution is cancelled by that of the neutral current. In order
to simplify the analysis one may try to replace $\chi$ by an
average value $\bar{\chi}(E_{\nu})$, e.g. defined by:
 \beq
 \bar{\chi}(E_{\nu})=\frac{1}{T_{max}(E_{\nu})}\int_0^{T_{max}(E_{\nu})}
 \chi(E_{\nu},T)~dT
 \label{chiav}
 \eeq
 Then surprisingly one finds $\bar{\chi}(E_{\nu})$ is independent
 of $E_{\nu}$ with a constant value of $0.42$. This is perhaps a
 rather high price one may have to pay for detecting the neutrino
 oscillations as proposed in this work.
\subsection{Modification of neutrino oscillation in a magnetic field due to the
magnetic moment}
 In the presence of a neutrino magnetic moment (see the Appendix for derivations)
 one finds that the electron neutrino
disappearance probability in the three generation model discussed
above with Majorana neutrinos takes the form:
\begin{eqnarray}
P(\nu_eL \rightarrow \nu_eL)\approx  1&-&\frac{\left[ \sin^2{ 2
\theta_1}^2 +sin^2{2 \theta} \cos^2{ 2 \theta_1} \right]
 \sin^2{ \pi \frac{L}{L_{21}}} \sqrt{1+ (\xi \frac{B}{1T})^2}}
{(1+\delta^2)^2}\\
\nonumber
 &-& \frac{4 \delta^2 \sin^2(\pi
\frac{L}{L_{32}})}{(1+\delta^2)^2} \label{maj1}
\end{eqnarray}
The parameter $\theta_1$ describes the $L-R$ mixing due to the
neutrino magnetic moment (see the Appendix).

In the case of Dirac neutrinos (see the Appendix) the above
equation becomes
\begin{equation}
P(\nu_eL \rightarrow \nu_eL)\approx  1-\frac{(\sin 2 \theta)^2
\sin^2{\pi
\frac{L}{L_{21}}}\cos^2{(\mu_{\nu}BL)}-\sin^2{(\mu_{\nu}BL})+ 4
\delta^2 \sin^2{\pi \frac{L}{L_{32}}}}{(1+\delta^2)^2}
\label{dir1}
\end{equation}
The parameter $\xi$ describes the mixing due to the neutrino
magnetic moment (see the Appendix) and has been  evaluated in a
magnetic field of strength $1T$, is given as follows:
$$\xi=\frac{4 \mu_{\nu} E_{\nu}~1T}{\delta m^2_{21}}$$

\begin{figure}
\hspace*{-0.0 cm}
\includegraphics[height=.4\textheight]{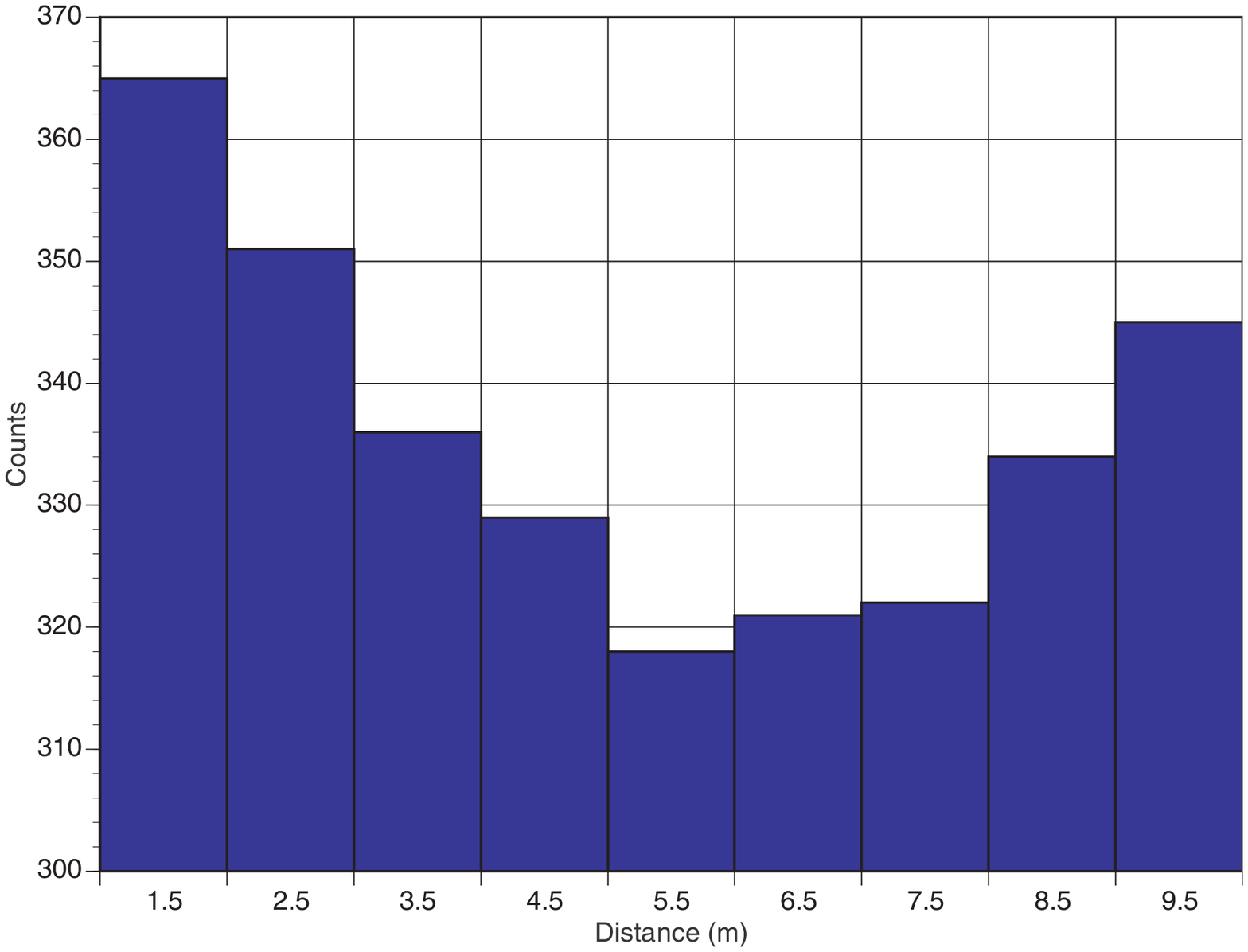}
\includegraphics[height=.4\textheight]{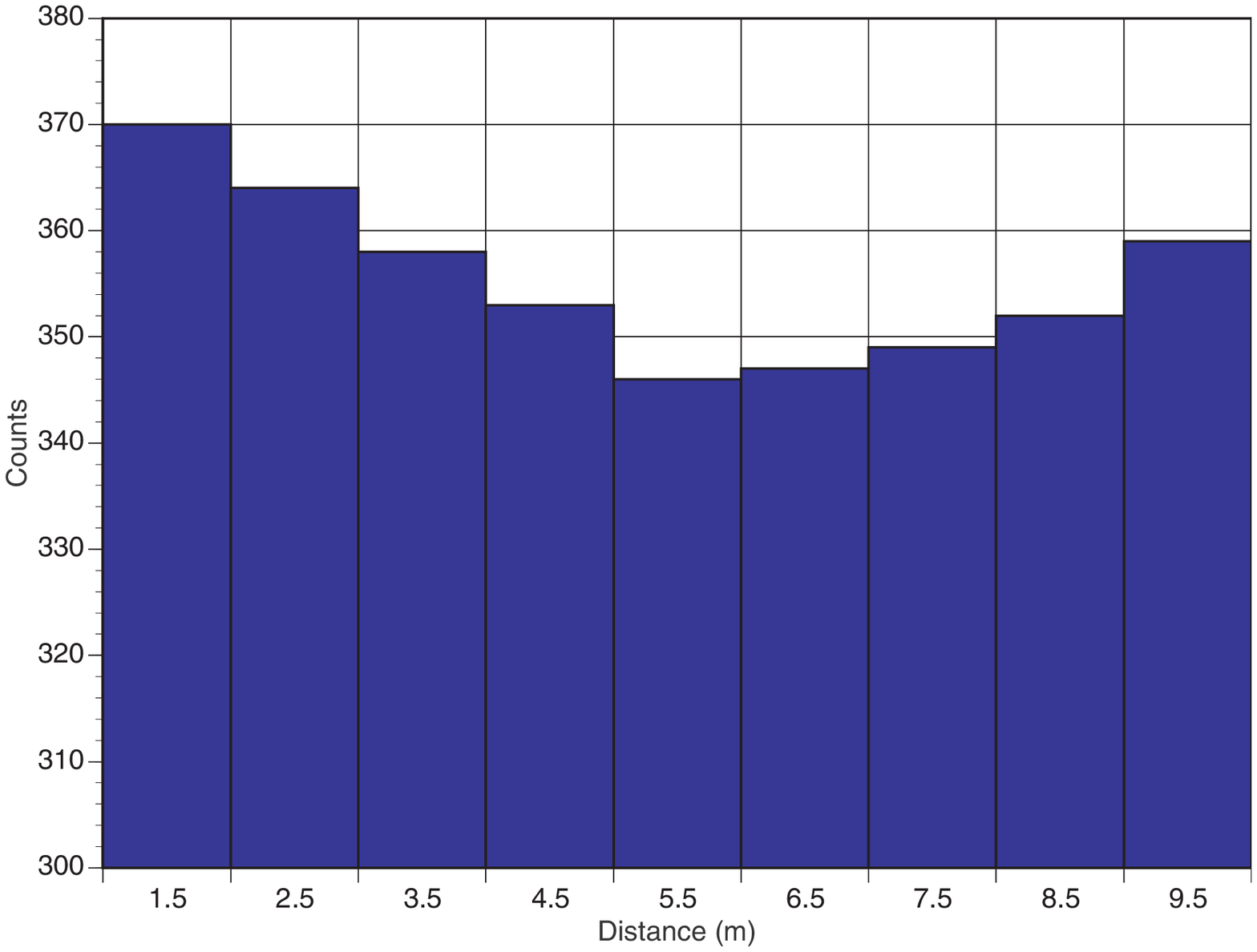}
\caption{ Simulation of $\nu_e$ disappearance due to the large
 $\Delta m^2=2.5 \times 10^{-3} (eV)^2$
involving the small mixing angle $\delta$. On the left we show
results for $\delta=0.170$, while on
 the right we show results for $\delta=0.085$. One expects to unambiguously see the
full oscillation inside the detector with the maximum
disappearance occurring around $5.5m$. This a bit smaller than the
calculated value of $7.5m$ due to the fact that the simulation
takes into account all energies. \label{osc} }
\end{figure}

\begin{figure}
\hspace*{-0.0 cm}
\includegraphics[height=.4\textheight]{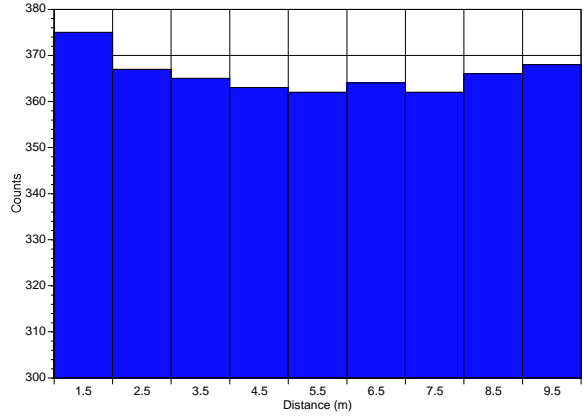}
\caption{ The same as in Fig. \ref{osc} for $\delta=0.045$.
\label{osc2} }
\end{figure}

 For $L_{12}>>L$, $L_{23} \approx L$ and for small magnetic
moment the above equations become:
\begin{equation}
P \left( \nu_{eL} \rightarrow \nu_{eL} \right) \approx 1-\frac{
\left[ \sin {2 \theta}^2 +\xi ^2 (B/1T)^2 \right](\pi
\frac{L}{L_{21}})^2 +4 \delta^2 \sin^2{ \pi
\frac{L}{L_{32}}}}{(1+\delta^2)^2}
 \label{dirmaj}
\end{equation}
We see that in the experiment involving a triton target one will
actually observe a sinusoidal oscillation as a function of the
source-detector distance $L$ with an amplitude, which is
proportional to the square of the small mixing angle $\delta$. The
relevant oscillation length is given by:
$$L_{32}=2.476m \frac{E_{\nu} (MeV)}{\Delta m^2_{32}((eV)^2)}$$
In the present experiment for an average neutrino energy
 $E_{\nu} \approx 13KeV$ and $\Delta m^2_{32}=2.5 \times 10^{-3} (eV)^2$ we
find
$$L_{32} \approx 13.5m$$
In other words the maximum will occur close to the source at about
$L=7.5m$. Simulations of the above neutrino oscillation involving
$\nu_e$ disappearance due to the large $\Delta m^2=2.5 \times
10^{-3}$, i.e associated with the small mixing $\delta$, are shown
in Figs \ref{osc}- \ref{osc2}. One clearly sees that the expected
oscillation, present even for $\delta$ as low as $0.045$, will
occur well inside the detector.

 Superimposed on this oscillation one will
see an effect due to the smaller mass difference, which will
increase quadratically with the distance $L$. In the presence of a
magnetic field this amplitude will also depend quadratically in
the magnetic field. The product of the mixing angle with the
mass-squared difference in one hand and the effect of the magnetic
moment squared on the other are additive. This is indeed very
beautiful experimental signature. Its practical exploitation
depends, of course, on the actual value of the parameter $\xi$.
This can be cast in the form:
\begin{equation}
 \xi=2 \times 10^{-10}
\frac{\mu_{\nu}}{10^{-12} \mu_B} \frac{E_{\nu}}{0.5MeV}
\frac{10^{-2}~eV^2} {\delta m^2_{21}} \label{xi.1}
\end{equation}
 It is unfortunately clear that the parameter $\xi$ is unobservable in
 all experiments involving low energy neutrinos,  if the magnetic
moment is of the order of $10^{-12} \mu_B$. This is particularly
true for the triton decay experiments with maximum energy of 18.6
KeV.

\subsection{Modification of neutrino oscillations due to neutrino decay}

 The effect of the finite life time  of  neutrinos on the
oscillation pattern \cite{VogEng} has mainly been investigated in
connection with solar neutrinos, see, e.g., Indumathi
\cite{INDUMATHI} and references therein. If in fact the decay
widths are very small only in the case of very long distances one
may have a detectable effect. In spite of this  we will examine
what effects, if any, the finite neutrino life time may have on
our experiment or other non-solar experiments. For the readers
convenience some useful formulas are given in the appendix. We
find that the decay width for the transition $i \rightarrow j$ is
given by the formula \cite{PAKVASA}:
 \beq
 \Gamma_i=K_0 \frac{m_i^2-m_j^2}{E_{\nu}}(1+\frac{m_j}{m_i})^{2}
 \label{decay5}
 \eeq
For more complete expressions, involving Majoron models
\cite{VERGADOS,MAJORON}, the reader is referred to the literature
\cite{BEACOM,RATES}. We prefer to use here the dimensionless
quantity $K_0$ rather the decay constant $\alpha=m_i/\tau_{0i}$.
These quantities are related via the equation $K_0=\alpha / \Delta
m^2$. A limit on the decay constant $\alpha$ exists from the SNO
data \cite{BEACOM}, $\alpha\leq10^4(m/s)=7 \times10^{-12}(eV)^2$,
but it is not very firm since depends on certain assumptions.
Anyway this leads to $K_0\leq 2.6 \times 10^{-9}$. For the
proposed experiment, $L\approx 10 m$ $<E_{\nu}>=10 KeV$, if we
demand $\Gamma L/2 \approx 3$, the dimensionless quantity $K_0$
has to be of order unity. For $K_0=0.5$ in the hierarchical case,
$ m_i^2 \approx \Delta m^2= m_i^2-m_j^2$, we obtain:
$$\Gamma_3=1.0 \times 10^{-7} eV~~~,~~~\Gamma_2=3.0 \times 10^{-9}
eV$$
 These are much larger than those obtained from
 the above bound and $\Gamma \approx10^{-26}eV\Leftrightarrow \tau=10^{11}s$
 or typical values expected in reasonable theoretical models \cite{VERGADOS}.
 In the case of solar neutrinos a value of $K_0 \approx 10^{-9}$ may be
 adequate, see Indumathi \cite{INDUMATHI}.

From the formulas in the appendix (Eqs. (\ref{decay.2} )-
(\ref{decay.4})) we  see that at sufficiently long distances the
neutrino oscillation is wiped out by the finite neutrino lifetime.
It is amusing to remark that for the rather unrealistic case
$\Gamma_3L /2 \approx 0.1$ the oscillation proportional to
$\delta^2$, the observation of which is one of the main goals of
this experiment, will be suppressed to $10\% $ of its value
without the presence of neutrino decay. In other words in this
case the extraction of the parameter $\delta$ may not be
straightforward.
\section{Elastic electron neutrino scattering.}
\label{secens}

 The elastic neutrino electron scattering is very crucial in our
 investigation, since it will be employed for the detection of
 neutrinos.

 We have seen in the previous section that the detection of the neutrino
 magnetic moment in laboratory neutrino oscillation experiments is extremely
difficult, if not impossible. The elastic electron neutrino
scattering offers a better possibility. Following the work of
Vogel and Engel \cite{VogEng} one can write the relevant
differential cross section as follows:
\begin{equation}
\frac{d\sigma}{dT}=\left(\frac{d\sigma}{dT}\right)_{week}+
\left(\frac{d\sigma}{dT}\right)_{EM} \label{elas1a}
\end{equation}
We ignored the contribution due to the neutrino charged radius.

The cross section due to weak interaction alone takes the form
\cite{VogEng}:
 \beq
 \left(\frac{d\sigma}{dT}\right)_{week}=\frac{G^2_F m_e}{2 \pi}
 \left[ (g_V+g_A)^2+ (g_V-g_A)^2 [1-\frac{T}{E_{\nu}}]^2+
 (g_A^2-g_V^2)\frac{m_eT}{E^2_{\nu}}     \right ]
 \label{elasw}
 \eeq
 where
 $$g_V=2\sin^2\theta_W+1/2~~for~~ \nu_e~~~~,~~~~g_V=2\sin^2\theta_W-1/2~~for~~ \nu_{\mu},\nu_{\tau}$$
 $$g_A=1/2~~for~~ \nu_e~~~~,~~~~g_A=-1/2~~for~~ \nu_{\mu},\nu_{\tau}$$
 For antineutrinos $g_A\rightarrow-g_A$. To set the scale we see
 that
\beq \frac{G^2_F m_e}{2 \pi}=0.445\times 10^{-48}~\frac{m^2}{MeV}
\label{weekval} \eeq
 In the above expressions for the $\nu_{\mu},\nu_{\tau}$ only the
 neutral current has been included, while for $\nu_e$ both the
 neutral and the charged current contribute.

 The second piece of the cross-section becomes:
\begin{equation}
\left(\frac{d\sigma}{dT}\right)_{EM}= \pi (\frac{ \alpha}{m_e})^2
(\frac{ \mu_{l}}{\mu_B})^2 \frac{1}{T} \left(1-\frac{T}{E_{\nu}}
\right) \label{elas1b}
\end{equation}
where
$$\mu^2_l=|\mu^2_{\nu}|$$
$$\mu^2_l=\left( \sin{(\alpha_{CP}/2)} \sin{2 \theta} \right)^2 |\mu_{\nu}|2$$
for Dirac and Majorana neutrinos respectively. The angle
$\alpha_{CP}$ is the relative CP phase of the Majorana neutrino
mass eigenstates. The contribution of the magnetic moment can also
be written as:
\begin{equation}
\left( \frac{d\sigma}{dT} \right)_{EM}=\sigma_0
 \left( \frac{\mu_l}{10^{-12}\mu_B} \right)^2 \frac{1}{T} \left( 1-\frac{T}{E_{\nu}} \right )
\label{elas2}
\end{equation}
The quantity $\sigma_0$ sets the scale for the cross section and
is quite small, $\sigma_0=2.5 \times 10^{-25}b$.

 The electron energy depends on the neutrino energy and the
scattering angle and is given by:
$$T=\frac{X^2}{2 m_e}~~,~~X=2E_{\nu} \frac{m_e(m_e+E_{\nu})\cos{\theta}}
{(m_e+E_\nu)^2-(E_{\nu} \cos{\theta})^2}$$
The last equation can be simplified as follows:
$$T \approx \frac{ 2(E_\nu \cos{\theta})^2}{m_e}$$
 The electron
energy depends on the neutrino spectrum. For $E_{\nu}=18.6~KeV$
one finds that the maximum electron kinetic energy approximately
is \cite{GIOMATAR}:
$$ T_{max}=1.27~KeV$$
 Integrating the differential cross section between $0.1$ and $1.27~KeV$ we find that the total
cross section is:
$$\sigma=2.5~\sigma_0$$
It is tempting for comparison to express the above EM differential
cross section in terms of the week interaction as follows:
\begin{equation}
\left( \frac{d\sigma}{dT} \right)_{EM}=\xi^2_1 \frac{2 G_F^2 m_e}
{\pi} \left( \frac{\mu_l}{10^{-12}\mu_B} \right)^2
\frac{1.27KeV}{T}  \left( 1-\frac{T}{E_{\nu}} \right )
\label{elas3}
\end{equation}
The parameter $\xi_1$ essentially gives the ratio of the
interaction due to the magnetic moment divided by that of the week
interaction. Evaluated  at the energy of $1.27 KeV$ it becomes:
$$\xi_1 \approx 0.39 $$
So the magnetic moment at these low energies can make a detectable
contribution provided that it is not much smaller than
 $10^{-12} \mu_B$.
 In many cases one would like to know the difference between the
 cross section of the electronic neutrino and that of one of the
 other flavors, i.e.
 \beq
 \chi(E_{\nu},T)=\frac{(d\sigma(\nu_e,e^-))/dT-d(\sigma(\nu_{\alpha},e^-))/dT}{d(\sigma(\nu_e,e^-))/dT}
 \label{elas4}
 \eeq
with $\nu_{\alpha}$  is either $\nu_{\mu}$ or $\nu_{\tau}$). Then
from the above expression for the differential cross-section one
finds:
\beq
 \chi=2\frac{2-(m_eT/E^2_{\nu})}{(1+2\sin^2\theta_W)^2/(2sin^2\theta_W)
 +2sin^2\theta_W(1-T/E_{\nu})^2-(1+2\sin^2\theta_W)(m_eT/E^2_{\nu})}
 \label{elas5}
 \eeq

 \section{Radiative neutrino decay}
 \label{secrnd}
 Using the formulas obtained in the appendix we can compute the
 differential and total radiative neutrino decay width. This is
 due to the neutrino magnetic moment.
 In a hierarchical scheme, $m_1<<m_2<<m_3$, and assuming that the
 neutrino masses are much smaller than the neutrino energy
  the differential width for the electron
neutrino decay
 $$\frac{d \Gamma_{\gamma}(\nu_e(E_{\nu}) \rightarrow
\nu_e (E_{\nu}-k) )}{d k}$$
 takes the form
\begin{eqnarray}
 \frac{d \Gamma_{\gamma}(\nu_e \rightarrow \nu_e)}{d k}&=&
 10^{-24} \frac{\alpha}{2}
\frac{1}{(2 m_e)^2  4 E_{\nu}}\\
\nonumber & [& \left(4
\delta^2\frac{|\mu_{31}|^2+|\mu_{32}|^2}{(10^{-12}\mu_B)^2} \right
) (\Delta m^2_{32})^{3/2}
 +\sin^2{ 2\theta_{solar}} \frac{|\mu_{21}|^2}{(10^{-12} \mu_B)^2} (\Delta m^2 _{21})^{3/2} ]
 \label{raterad.3}
\end{eqnarray}
The range of the photon energy (see the appendix) is
$k_{min}=(1-\frac{m_i^2}{m_j^2})\frac{E_i-p_i}{2}\approx
\frac{m^2_i}{4p_i}\approx 0$,
$k_{max}=(1-\frac{m_i^2}{m_j^2})\frac{E_i+p_i}{2}\approx
p_i\approx E_{\nu}$ i.e. $0 \leq k \leq E_{\nu}$.
Thus the total radiative decay width   takes the form:
\begin{eqnarray}
  \Gamma_{\gamma}(\nu_e\rightarrow \nu_e )&=&
  10^{-24} \frac{\alpha}{2}
\frac{1}{(2 m_e)^2}\frac{1}{4}\\
 \nonumber
& [& \left(4
\delta^2\frac{|\mu_{31}|^2+|\mu_{32}|^2}{(10^{-12}\mu_B)^2} \right
) (\Delta m^2_{32})^{3/2}
 +\sin^2{ 2\theta_{solar}} \frac{|\mu_{21}|^2}{(10^{-12} \mu_B)^2} (\Delta m^2 _{21})^{3/2} ]
 \label{raterad.4}
 \end{eqnarray}
 With the above approximations  the differential rate can be cast in the form:
 \beq
\frac{d \Gamma_{\gamma}(\nu_e \rightarrow \nu_e)}{d k}=\frac{
\Gamma_{\gamma}(\nu_e \rightarrow \nu_e)}{E_{\nu}}
\label{raterad.5} \eeq We notice that only the off diagonal
elements of the magnetic moment appear, reminiscent of the fact
that a charged particle cannot radiate off photons in the absence
of matter, i.e in the absence of an external electromagnetic
field. This means that radiative decay will be  further
suppressed, if the neutrinos are Dirac particles, a GIM like
effect. No such suppression occurs in the Majorana neutrino case,
since the diagonal elements of the magnetic moment are zero
anyway. We should also point out that, since for real photons we
have one electromagnetic interaction less, the rate contains one
power of $\alpha$ less compared to that associated with
neutrino-electron scattering.

 The width due to the small mixing $\delta$ for
$\mu_{32}=10^{-12}\mu_B$ is $\Gamma_{\gamma}(\nu_e\rightarrow
\nu_e )=4.5\times10^{-45}eV\Longleftrightarrow\tau=1.7\times
10^{11}s$. Again for $\mu_{21}=10^{-12}\mu_B$
 the width due to the $\sin^2{2\theta}$ is even smaller,
$\Gamma_{\gamma}(\nu_e\rightarrow \nu_e
)=3.5\times10^{-47}eV\Longleftrightarrow\tau=2.2\times 10^{13}s$,
since the advantage of the larger mixing angle is lost due to the
smaller neutrino mass squared difference. The above lifetimes are
on the short side when compared to model calculations
\cite{VERGADOS}. In spite of this the above radiative  widths seem
too small for their observation in the case of decay of
terrestrial neutrinos, but reasonable for solar neutrinos or other
astrophysical observations.
\section{Radiative neutrino-electron scattering}
\label{secrens}
\begin{figure} \hspace*{-0.0 cm}
\includegraphics[height=.4\textheight]{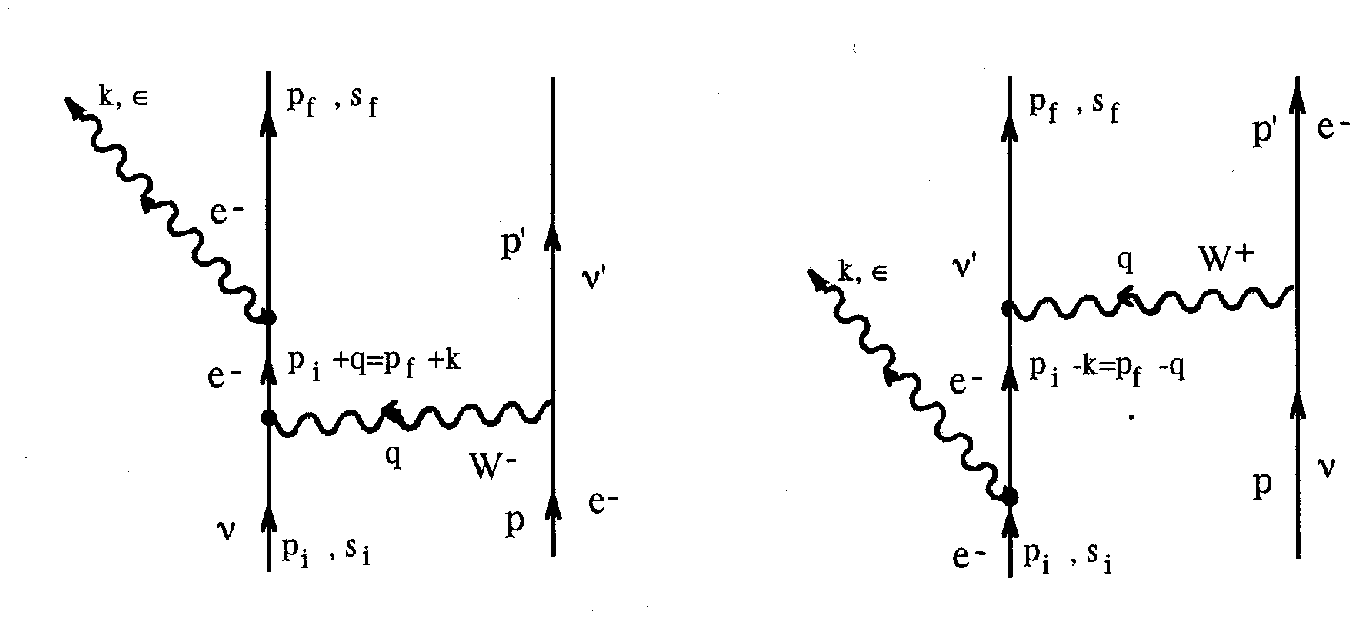}
\caption{ The Feynman diagrams contributing to radiative neutrino
electron scattering via the charged current. \label{charged} }
\end{figure}

\begin{figure}
\hspace*{-0.0 cm}
\includegraphics[height=.4\textheight]{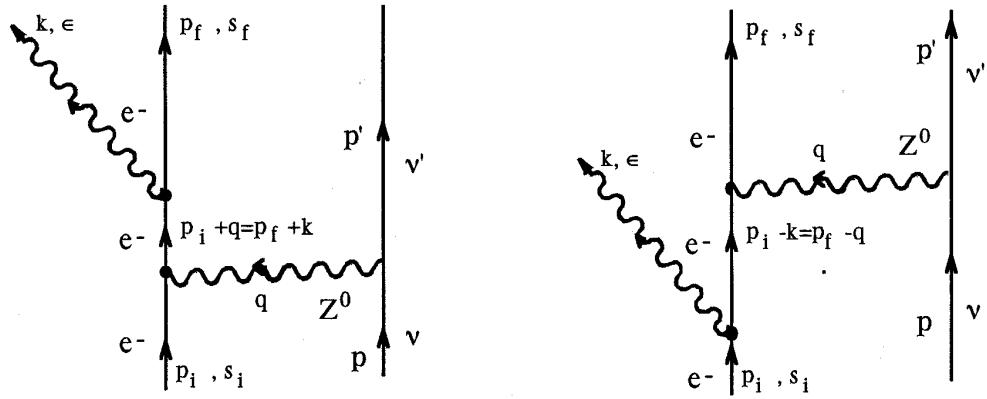}
\caption{ The Feynman diagrams contributing to radiative neutrino
electron scattering via the neutral current. \label{neutral} }
\end{figure}

 We have seen in the previous section that the
radiative neutrino decay for low energy neutrinos is perhaps
unobservable. In this section we will examine the radiative
neutrino decay in the presence of matter, in our case electrons.
$$\nu_e(p_{\nu})+e^- \longrightarrow \nu_e (p^{'}_{\nu})+e^-(p_e)+\gamma(k)$$
 This occurs  via the collaborative
effect of electromagnetic and week interactions as is shown in
Fig. \ref{charged} and \ref{neutral}, for the charged and neutral
current respectively.

 The evaluation of the cross section
associated with these diagrams is rather complicated, but in the
present case the electrons are extremely non relativistic. Thus in
the inverse of the (intermediate) electron propagator we can
retain the mass rather than the momenta (exact results without
this approximation will appear elsewhere). Then after some
tedious, but straight-forward, trace calculations one can perform
the angular integrals over the three-body final states to get:
 \beq
k\frac{d\sigma(k,p_e)}{dk}=\sigma_{\gamma}\frac{p_edp_e}{m_e^2}
[\rho+(1+\rho)\frac{p_e^2}{4E^2_{\nu}}]
\label{radia1} \eeq where
$$\sigma_{\gamma}=\frac{37}{2\pi^2}(G_F~m_e)^2~\alpha
\approx6.8\times 10^{-14}pb$$ sets the scale for this process,
$\rho=5.4\times10^{-2}$ is the ratio of the neutral to charged
current contribution and $p_e$ is the final electron momentum.
This momentum depends on the photon momentum k and the scattering
angles. For a given $k$  is restricted as follows:
$$0 \leq p_e \leq E_{\nu}-k$$
with maximum electron energy given by:
$$T_e = \frac{2(E_{\nu}-k)^2}{m_e}$$
From the above equations we cam immediately see that this process
is roughly of order $\alpha$ down compared to the week
neutrino-electron scattering cross-section. We also notice that
the total cross section diverges logarithmically as the photon
momentum goes to zero, reminiscent of the infrared divergence of
Bremsstrahlung radiation. In our case we will adopt a lower photon
momentum cutoff as imposed by our detector. We also notice that
$\sigma_{\gamma}$, characterizing  this process, is only a factor
of three smaller than $\sigma_0$ characterizing the neutrino
electron scattering cross section due to the magnetic moment. We
should bare in mind, however, that:
\begin{enumerate}
\item The  magnetic moment is not known. $\sigma_0$ was obtained
with the rather optimistic value $\mu_{\nu}=10^{-12}\mu_{B}$,
which is two orders of magnitude smaller than the present
experimental limit.
\item One now has the advantage of observing not only the electron but the photon as
well.
 \end{enumerate}

Integrating over the electron momenta we get
 \beq
k\frac{d\sigma(k}{dk}=\sigma_{\gamma}~2
\left[\frac{E_{\nu}-k}{m_e}\right]^2 [\rho
+2(1+\rho)\frac{(E_{\nu}-k)^2}{E^2_{\nu}}]
 \label{radia2}
\eeq
 Integrating this cross-section with respect
to the photon momentum we get:
 \beq
\sigma_{total}=\sigma_{\gamma}~2\frac{E_{\nu}^2}{m^2_e}[(2+3
\rho)\ell n\frac{E_{\nu}}{E_{cutoff}}-\frac{4+17 \rho}{6}]
 \label{radia3}
  \eeq
  with the energy cutoff such  lower than that required to make the
   above expression in the
 brackets vanish. Notice that the total cross-section is
 reduced further from the value $\sigma_{\gamma}$ by the ratio of
 the square of the neutrino energy divided by the electron mass.

 We have considered in our discussion only electron targets. For such low
 energy neutrinos the charged current cannot operate on hadronic targets,
  since this process is not allowed so long as the target, being stable,
  is not capable of undergoing ordinary $\beta$  decay. The
  neutral current, however, can always make a contribution.

\section {Summary and outlook}

The perspective of the experiment is to provide high statistics
-redundant, high precision measurement and minimize as much as
possible the systematic uncertainties of experimental origin,
which could be the main worry in the results of existing
experiments. The physics goals of the new atmospheric neutrino
measurement are summarized as follows:
\begin{enumerate}
\item Establish the phenomenon of neutrino oscillations with a different
Experimental technique free of systematic biases. Thus one hopes
to measure all the oscillation parameters, including the small
mixing angle in the electronic neutrino, clarifying this way the
nature of the oscillation mechanism.
\item A high sensitivity measurement  of the neutrino magnetic
moment, via electron neutrino scattering. At the same time
radiative electron neutrino scattering will be investigated,
exploiting the additional photon signature.
\item A new experimental investigation of neutrino decay.
\item Other novel
improvements of the experimental sensitivity are possible and must
be investigated. The benefit of increasing the gas pressure of the
detector that increases proportionally the number of events must
be investigated. A significant increase of event rate is a great
step forward improving the experimental accuracy and reducing the
impact of background rates.
\end{enumerate}

\section{Appendix}
\label{appen}
\subsection{Modification of neutrino
oscillations due to the magnetic moment}

The electromagnetic interaction between the mass neutrino
eigenstates $\nu_i$ and $\nu_j$ due to the magnetic moment
$\mu_{ji}$ takes the form:
$$\textit{L}_{EM}=\mu_{ji}\bar{u}(p_j)
\sigma_{\lambda \rho}q^{\rho}u(p_i)\epsilon^{\lambda}~,~\mu_{ji}=
10^{-12} \frac{\sqrt{4\pi\alpha}}{2 m_e}(\frac{\mu_{ji}}{10^{-12}\mu_B})$$
 with $\epsilon^{\lambda}$ the photon polarization and
$$\sigma_{\lambda\rho}=\frac{1}{2i} \left [
\gamma_{\lambda},\gamma_{\rho} \right ]~~,~~q=p_i-p_j~~$$
 For the moment  will limit ourselves  to the case of two generations only, i.e.
we will ignore the small mixing $\delta$. The two  cases of Dirac
and Majorana neutrinos must be treated separately.
\subsubsection{Majorana Neutrinos.}

 The  flavor states are related to the  mass eigenstates as follows:
$$ \nu_{eL}=\cos {\theta} \nu_{1L} + \sin{\theta} \nu_{2L} $$
$$ \nu_{\mu L}=-\sin {\theta} \nu_{1L} + \cos{\theta} \nu_{2L} $$
$$ \nu_{eR}^C=\cos {\theta} \nu_{1R} + \sin{\theta} e^{i\alpha} \nu_{2R} $$
$$ \nu_{\mu R}^C=-\sin {\theta} \nu_{1L} + \cos{\theta} e^{i\alpha} \nu_{2L} $$
In the above expressions for convenience we dropped the subscript
"solar" from the mixing angle.
 The above results are modified if the neutrinos have a magnetic moment and are found
in a magnetic field. Then the left and the right neutrino fields
are coupled via the dipole magnetic transition $\mu_{12}=\mu_{\nu}$. The
diagonal elements of the magnetic transition operator (magnetic moments of the
mass eigenstates) are zero (the Majorana
neutrinos do not possess electromagnetic properties). Furthermore
the magnetic moment submatrices are antisymmetric. Thus the
 mass eigenstates evolve as follows:

$\left ( \begin{array}{c} \frac{d\nu_{1L}}{dt} \\
  \frac{d\nu_{2L}}{dt} \\
 \frac{d\nu_{1R}}{dt} \\
 \frac{d\nu_{2R}}{dt} \\
\end{array} \right )=
\left ( \begin{array}{cccc}
E_1&0&0&\mu_{\nu}B\\
0&E_2&-\mu_{\nu}B&0\\
0&-\mu_{\nu}B&E_1&0\\
\mu_{\nu}B&0&0&E_2\\
 \end{array} \right )=
\left ( \begin{array}{c}\nu_{1L} \\
\nu_{2L} \\
\nu_{1R} \\
\nu_{2R}
\end{array} \right )$\\
The eigenvalues of the above matrix are:
$$\lambda_{\pm}=\frac{E_1+E_2 \pm \sqrt{(E_1-E_2)^2+4(\mu_{\nu}B)^2}}{2}$$
(doubly degenerate), while the eigenvectors, indicated via
 $\eta_1^+,\eta_2^+,eta_1^-,\eta_2^-$, are related to the mass eigenstates as follows:
$$\nu_{1L}=\sin{\theta_1}\eta_1^{+}+\cos{\theta_1} \eta_1^{-}$$
$$\nu_{2L}=\cos{\theta_1}\eta_2^{+}+\sin{\theta_1} \eta_2^{-}$$
$$\nu_{1R}=-\sin{\theta_1}\eta_2^{+}+\cos{\theta_1} \eta_2^{-}$$
$$\nu_{2R}=\cos{\theta_1}\eta_1^{+}-\sin{\theta_1} \eta_1^{-}$$
with the mixing angle defined by:
$$ \tan{2 \theta_1}=\frac{2\mu_{\nu}B}{E_2-E_1}=
\frac{4\mu_{\nu}B E_{\nu}}{\Delta m _{21}^2}$$ Restricting oneself
to two generations one now has the following pattern
 of neutrino oscillations:
$$P\left( \nu_{eL} \rightarrow \nu_{\mu L} \right)=\left[ \cos{2 \theta _1}
 \sin{2 \theta} \sin{ \pi \frac{L}{L_{21}} \sqrt {1+ (\xi \frac{B}{1T})^2}} \right]^2$$
$$P\left(\nu_{eL} \rightarrow \nu^C_{eR} \right)= \sin^2{(\alpha/2)}
\sin^2{2 \theta_ 1} \left[ \sin{ \pi \frac{L}{L_{21}}\sqrt{1+ (\xi
\frac {B}{1T})^2}} \right] ^2$$
$$P\left(\nu_{eL} \rightarrow \nu ^C_{\mu R} \right)=
\sin^2{2 \theta_1} \sin^2{(\alpha/2)}
 \left[1-\sin{2 \theta}
\sin{ \pi \frac{L}{L_{21}}\sqrt{ 1+ (\xi \frac{B}{1T})^2}} \right]
^2$$
$$P\left(\nu_{eL} \rightarrow \nu_{eL} \right)=1-\left[ \sin{2 \theta_1}
 +cos^2{2 \theta_1}\sin^2{2 \theta} \right ]
 \sin^2{ \pi \frac{L}{L_{21}}\sqrt{1+ (\xi \frac{B}{1T})^2}}$$
with the parameter $\xi$ describes the mixing due to the neutrino
magnetic moment, in a magnetic field taken to be of strength $1T$,
and is given as follows:
$$\xi=\frac{4 \mu_{\nu} E_{\nu}~1T}{\delta m^2_{21}}$$
We see that the oscillation
 $P\left( \nu_{eL} \rightarrow \nu^C_{eR} \right)$
  vanishes when the relative $CP$ phase of the
two eigenstates, $\alpha$, is zero, since, then, the flavor
neutrinos are themselves Majorana particles.
 The electron neutrino
disappearance probability in the three generation model discussed
above takes the form:
\begin{eqnarray}
P(\nu_eL \rightarrow \nu_eL)\approx  1&-&\frac{\left[ \sin^2{ 2
\theta_1}^2 +sin^2{2 \theta} \cos^2{ 2 \theta_1} \right]
 \sin^2{ \pi \frac{L}{L_{21}}} \sqrt{1+ (\xi \frac{B}{1T})^2}}
{(1+\delta^2)^2}\\
\nonumber
 &-& \frac{4 \delta^2 \sin^2(\pi
\frac{L}{L_{32}})}{(1+\delta^2)^2} \label{dismaj1}
\end{eqnarray}
We notice that the disappearance probability is independent of the
phase $\alpha$. For $L_{12}>>L$, $L_{23} \approx L$ and for small
magnetic moment the above equation becomes:
\begin{equation}
P \left( \nu_{eL} \rightarrow \nu_{eL} \right) \approx 1-\frac{
\left[ \sin {2 \theta}^2 +\xi ^2 (B/1T)^2 \right](\pi
\frac{L}{L_{21}})^2 +4 \delta^2 \sin^2{ \pi
\frac{L}{L_{32}}}}{(1+\delta^2)^2}
 \label{dismaj2}
\end{equation}
We see that in the experiment involving a triton target one will
actually observe a sinusoidal oscillation as a function of the
source-detector distance $L$ with an amplitude, which is
proportional to the square of the small mixing angle $\delta$. The
relevant oscillation length is given by:
$$L_{32}=2.476m \frac{E_{\nu} (MeV)}{\Delta m^2_{32}((eV)^2)}$$
In the present experiment for an average neutrino energy
 $E_{\nu} \approx 15KeV$ and $\Delta m^2_{32}=2.5 \times 10^{-3} (eV)^2$ we
find
$$L_{32} \approx 15m$$
In other words the maximum will occur close to the source at about
$L=7.5m$

 Superimposed on this oscillation one will
see an effect due to the smaller mass difference, which will
increase quadratically with the distance $L$. In the presence of a
magnetic field this will also depend quadratically in the magnetic
field. The product of the mixing angle with the mass-squared
difference in one hand and the effect of the magnetic moment
squared  are additive. This is indeed very beautiful experimental
signature. Its practical exploitation depends, of course on the
actual value of the parameter $\xi$. This can be cast in the form:
\begin{equation}
 \xi=2 \times 10^{-10}
\frac{\mu_{\nu}}{10^{-12} \mu_B} \frac{E_{\nu}}{0.5MeV}
\frac{10^{-2}~eV^2} {\delta m^2_{21}} \label{xi}
\end{equation}
 It is unfortunately clear that the parameter $\xi$ is unobservable in reactor
experiments involving low energy neutrinos, even if the magnetic
moment of the order of $10^{-12} \mu_B$. This is particularly true
for the triton decay experiments with maximum energy of 18.6 KeV.
\subsubsection{Dirac Neutrinos.}
 The case of Dirac neutrinos is not favored in most gauge theories. It is, however, a
possibility favored in recent brane theories in extra dimensions.
The  flavor states are now related to the  mass eigenstates as
follows:
$$ \nu_{eL}=\cos {\theta} \nu_{1L} + \sin{\theta} \nu_{2L} $$
$$ \nu_{\mu L}=-\sin {\theta} \nu_{1L} + \cos{\theta} \nu_{2L} $$
$$ \nu_{eR}=\cos {\theta} \nu_{1L} + \sin{\theta}  \nu_{2L} $$
$$ \nu_{\mu R}=-\sin {\theta} \nu_{1R} + \cos{\theta} e^{i\alpha} \nu_{2R} $$
 The above results are also modified if the neutrinos have a magnetic moment
 and are found
in a magnetic field. We will assume that, in the mass eigenstate
basis, the off
 diagonal masses are much smaller than
the off diagonal ones (GIM effect).We will consider the case of
the inverse
 hierarchical masses, in which
 case in the masses $m_1$ and$m_2$ are almost degenerate. As a result the magnetic
 moments, which are proportional to the corresponding masses, are also almost equal.
In this case the mass eigenstates evolve as follows:\\
$\left ( \begin{array}{c} \frac{d\nu_{1L}}{dt} \\
  \frac{d\nu_{2L}}{dt} \\
\frac{d\nu_{1R}}{dt} \\
 \frac{d\nu_{2R}}{dt} \\
\end{array} \right )=
\left ( \begin{array}{cccc}
E_1&0&\mu_{\nu}B&0\\
0&E_2&0&\mu_{\nu}B\\
\mu_{\nu}B&0&E_1&0\\
0&\mu_{\nu}B&0&E_2\\
 \end{array} \right )=
\left ( \begin{array}{c}
\nu_{1L} \\
\nu_{2L} \\
\nu_{1R} \\
\nu_{2R}
\end{array} \right )$\\
The eigenvalues of the above matrix are:
$$\lambda_1=E_1+\mu_{\nu} B,\lambda_2=E_1-\mu_{\nu} B,\lambda_3=E_2+\mu_{\nu} B,
\lambda_4=E_2-\mu_{\nu} B,$$
 while the eigenvectors, indicated via
 $\eta_1,\eta_2,\eta_3,\eta_4 $, are related to the mass eigenstates as follows:
$$\nu_{1L}=\frac{1}{\sqrt 2}(\eta_1+ \eta_2)$$
$$\nu_{2L}=\frac{1}{\sqrt 2}(\eta_3+ \eta_4)$$
$$\nu_{1R}=\frac{1}{\sqrt 2}(\eta_1- \eta_2)$$
$$\nu_{2R}=\frac{1}{\sqrt 2}(\eta_3-\eta_4)$$
The following pattern of neutrino oscillations emerges:
$$P\left(\nu_{eL} \rightarrow \nu_{\mu L} \right)=
\left[ \cos{(\mu_{\nu}BL)} \sin{2 \theta}
 \sin{ \pi \frac{L}{L_{21}}} \right] ^2$$
$$P\left(\nu_{eL} \rightarrow \nu_{eR} \right)= \sin^2{(\mu_{\nu}BL)} \left[1-
\sin{2 \theta} \sin{ \pi \frac{L}{L_{21}}} \right] ^2$$
$$P\left(\nu_{eL} \rightarrow \nu_{\mu R} \right)=\left[ \sin{(\mu_{\nu}BL)} \sin{2 \theta}
 \sin{ \pi \frac{L}{L_{21}}} \right] ^2$$
$$P\left(\nu_{eL} \rightarrow \nu_{eL} \right)= \cos^2{(\mu_{\nu}BL)} \left[ 1-
\sin{2 \theta} \sin{ \pi \frac{L}{L_{21}}} \right] ^2$$
 The three generation electron neutrino disappearance probability takes the
form:
\begin{equation}
P(\nu_eL \rightarrow \nu_eL)\approx  1-\frac{(\sin 2 \theta)^2
\sin^2{\pi
\frac{L}{L_{21}}}\cos^2{(\mu_{\nu}BL)}-\sin^2{\mu_{(\nu}BL})+ 4
\delta^2 \sin^2{\pi \frac{L}{L_{32}}}}{(1+\delta^2)^2}
\label{disdir1}
\end{equation}
For $L_{12}>>L$, $L_{23} \approx L$ and for small magnetic moment
the above equation becomes:
\begin{equation}
P \left( \nu_{eL} \rightarrow \nu_{eL} \right)\approx
1-\frac{\left[ \sin{ 2 \theta}^2 +\xi^2 (B/1T)^2 \right]
(\pi\frac{L}{L_{21}})^2 + 4 \delta^2 \sin^2{\pi
\frac{L}{L_{32}}}}{(1+\delta^2)^2} \label{disdir2}
\end{equation}
We see that in this limit one cannot distinguish between the Dirac
and Majorana neutrino oscillation  patterns (compare eqs
(\ref{dismaj2}) and (\ref{disdir2})).

\subsection{Modification of neutrino oscillations due to decay}

We will suppose the normal hierarchy of neutrino masses. In the
presence of a finite neutrino decay width the quantum evolution of
the neutrino
  mass eigenstate takes the form:
 \beq
\nu_i(t)=\nu_i(0)Exp[-(iE_i +\Gamma _i/2)t]
 \label{decay.1}
 \eeq
 where $\Gamma_i,~i=2,3$ is the decay, width with respect to the
 laboratory, of $\nu_i$. The neutrino $\nu_1$ is assumed to be absolutely
  stable. The most common decay modes are \cite{VERGADOS}:
 \beq
 \nu_{iL} \rightarrow \nu_{jL}+\gamma
 \label{life.1}
 \eeq
 radiative decay, which is expected to be quite slow. Or
\beq
 \nu_i \rightarrow \nu_j+\nu_k+\bar{\nu}_k
 \label{life.2}
 \eeq
 which is also expected to be slow. One may also have
\beq
 \nu_{iL} \rightarrow \nu_{jR}+\phi^0
 \label{life.3}
 \eeq
which requires the existence of a very light physical Higgs scalar
and sterile neutrinos. The most dominant mechanism is perhaps the
Majoron emission:
 \beq
 \nu_{iL} \rightarrow \bar{\nu}_{jR}+\chi^0
 \label{life.}
 \eeq
 available only to  Majorana neutrinos.

 In all the above channels the final neutrino $\nu_j$ is different from
 $\nu_1$ (in the case of processes (\ref{life.1}) and (\ref{life.2}) the final
 neutrino has a different energy, so it cannot be mistaken with
 $\nu_1$)

 The widths $\Gamma_k=1/ \tau_k$ can be related to those in the rest frame
 $\Gamma_{0k}=1/ \tau_{0k}$ using the well known relation
 $$\tau_k=\gamma \tau_{0k}=(E_{\nu_e}/m_k)\tau_{0k}$$
 The life-time in the rest frame can be cast in the form
 \cite{PAKVASA}
 $$\tau_0=\frac{1}{K_0}\frac{m_i}{m_i^2-m_j^2}(1+\frac{m_j}{m_i})^{-2}$$
 We thus get
 \beq
 \Gamma_i=K_0 \frac{m_i^2-m_j^2}{E_{\nu}}(1+\frac{m_j}{m_i})^{2}
 \label{decay.5}
 \eeq
 The parameter $K_0$ depends on the model. For a lifetime for
 $\nu_3$ of the order of $10^{11}s$ we need $K_0$ of order of $
 10^{-10}$.

 Since the neutrino $\nu_1$ can neither decay nor be repopulated, the relevant
 amplitude entering $\nu_e$ disappearance takes the form:
 $$\cos^2{\theta} Exp[-iE_1]+\sin^2{\theta} Exp[-(iE_2 +\Gamma
 _2/2)t]+\delta^2 Exp[-(iE_3 +\Gamma _3/2)t]$$
  Thus the disappearance probability can be written as:
 \beq
 P(\nu_eL \rightarrow \nu_eL)=A^2(L)+B(L)-
\frac{\sin^2 2\theta \sin^2(\pi \frac{L}{L_{21}})+4 \delta^2
\sin^2(\pi \frac{L}{L_{32}})}{(1+\delta^2)^2}
 \label{decay.2}
 \eeq
 with $A(L)$ describing the part due to decay alone and $B(L)$
 providing the collaborative effect due to oscillation and decay.
 One finds:
 \beq
 A(L)=1-\frac{\sin^2{\theta}[1-Exp(-\Gamma_2L/2)]+\delta^2[1-Exp(-\Gamma_3L/2)]}
 {1+\delta^2}
 \label{decay.3}
 \eeq
\begin{eqnarray}
 B(L)&=& \frac{\sin 2 \theta^2 \sin^2(\pi
\frac{L}{L_{21}})[1-Exp(-\Gamma_2L/2)]}{(1+\delta^2)^2}\\
\nonumber
 &+& \frac{4 \delta^2 \sin^2(\pi \frac{L}{L_{32}}) \left
[1-Exp(-\Gamma_3L/2)~[\cos^2{\theta}+\sin^2{\theta}
Exp(-\Gamma_2L/2)]\ \right ]}{(1+\delta^2)^2} \label{decay.4}
\end{eqnarray}
 In the absence of neutrino oscillations the disappearance is
 given by the function $A(L)$, which has the properties:
 $$B(0)=0~~,~~B(\infty)= \frac{\sin^2 2\theta \sin^2(\pi \frac{L}{L_{21}})+4 \delta^2
 \sin^2(\pi\frac{L}{L_{32}})}{(1+\delta^2)}$$
 $$A(0)=1~~,~~A(\infty)=\frac{\cos^4{\theta_{solar}}}{(1+\delta^2)^2}$$
In other words the neutrino oscillation is wiped out at sufficiently long distances.
\subsection{Derivation of the neutrino decay width}
We are now going to study the radiative neutrino decay, see
(\ref{life.1}), a bit further. The differential decay width for a
neutrino $\nu_i$ with mass $m_i$ to a neutrino $\nu_j$ with mass
$m_j$, $m_i>m_j$, with production of a photon of momentum $k$ is
given by:
 \beq
 \frac{d \Gamma(i \rightarrow j)}{d k}=10^{-24}
\frac{\alpha}{2} \frac{1}{(2 m_e)^2 p_i} \left |\frac
{\mu_{ij}}{10^{-12} \mu_B} \right|^2 \frac{ \left( m_i+m_j \right
)^4} {m_i}
 \label{rad.1}
\eeq
where $p_i$ is the momentum of the initial neutrino and the
range of the photon momentum k is given by:
$$ (1-\frac{m^2_j}{m^2_i}) \frac{E_i-p_i}{2} \leq k \leq (1-\frac{m^2_j}{m^2_i})\frac{E_i+p_i}{2}$$
The total rate takes the form:
 \beq
 \Gamma(i \rightarrow j)=10^{-24}
\frac{\alpha}{2}\frac{1}{ (2 m_e)^2}\left |\frac
{\mu_{ij}}{10^{-12}\mu_B} \right|^2 (1-\frac{m^2_j}{m^2_i})\frac{
\left( m_i+m_j \right )^4} {m_i} \label{rad.2} \eeq
 The differential rate for electron
neutrino disappearance takes the form \beq \frac{d \Gamma
(\nu_e(E_{\nu}) \rightarrow \nu_e (E_{\nu}-k) )}{d k}=10^{-24}
\sum_{j <i } \frac{\alpha}{2 }\frac{1}{(2 m_e)^2 2 p_i}\left
|\Lambda_{ij} \right|^2 \frac{ \left( m_i+m_j \right)^4} {m_i}
\label{rad.3} \eeq while the total rate takes the form: \beq
\Gamma(\nu_e \rightarrow \nu_e )= 10^{-24} \sum_{j <i }
\frac{\alpha}{2 }\frac{1}{ (2 m _e)^2} \left |\Lambda_{ij}
\right|^2 \frac{ \left( m_i+m_j\right)^4} {m_i}
(1-\frac{m^2_j}{m^2_i}) \label{rad.4} \eeq with
$$\Lambda_{ij}=U^{*}_{ej}U_{ei}\frac{\mu_{ij}}{10^{-12} \mu_B}$$
with the mass hierarchy $m_j < m_i$

\end{document}